\documentclass[fleqn,twoside]{article}
\usepackage{espcrc2}

\usepackage{graphicx}

\newcommand{\AmS}{{\protect\the\textfont2
  A\kern-.1667em\lower.5ex\hbox{M}\kern-.125emS}}

\title{DLCQ Strings and Branched Covers of Torii} \author{Gordon
W. Semenoff\address[MCSD]{Department of Physics and Astronomy,
University of British Columbia, Vancouver, British Columbia, Canada
V6T 1Z1.\\~~\\~~\\~~Presented at the Light-Cone Workshop\\ "Light-cone
Physics: Particles and Strings" \\Trento, Italy, September, 2001}
\thanks{This work is supported in part by NSERC of Canada.}}
                
\begin{document}

\begin{abstract}
In this lecture I will review some results about the discrete
light-cone quantization (DLCQ) of strings and some connections of the
results with matrix string theory.  I will review arguments which show
that, in the path integral representation of the thermal free energy
of a string, the compactifications which are necessary to obtain
discrete light-cone quantization constrains the integral over all
Riemann surfaces of a given genus to the set of those Riemann surfaces
which are branched covers of a particular torus.  I then review an
explicit check of this result at genus 1. I discuss the intriguing
suggestion that these branched covers of a torus are related to those
which are found in a certain limit of the matrix string model.
\vspace{1pc}
\end{abstract}

\maketitle

\section{Preamble}

In this lecture I will review some aspects of the discrete light-cone
quantization (DLCQ) of strings.  This is a summary of work which has
appeared in a series of publications about this subject, its
relationship with the Matrix model of M-theory and related issues,
\cite{Grignani:1999sp}-\cite{Grignani:2001ik}.

One of the essential points will be that, when the light-cone is
compactified, the degrees of freedom of the string are thinned
somewhat. We will see this by examining the thermodynamic partition
function.  The thinning of degrees of freedom can be quantified
geometrically and summarized in a simple statement about string
worldsheets: The set of string worldsheets in DLCQ is the set of those
Riemann surfaces which are branched covers of a particular torus,
rather than the bigger set of all Riemann surfaces. This means that
the string fluctuates less.  A further lesson is that these branched
covers appear naturally in certain models of random matrices
\cite{Kostov:1997bn}-\cite{D'Adda:2001nf}.  The example that we are
particularly interested in here is matrix string theory where the
limit of weak string coupling of the model is related to string
degrees of freedom living on branched covers.

After discussing the general result, we will illustrate by doing three
different computations of the same quantity: the thermodynamic
partition function of discrete light-cone quantized strings.  The
first computation uses the path Polyakov path integral.  The second
computation takes the operator quantization of the free string and
solves for the spectrum and then constructs the thermodynamic
partition function from that spectrum.  The third computation begins
with a certain limit of the matrix model of M-theory.  In this limit
the model reduces to a statistical theory of eigenvalues of the
matrices which live on branched covers of a torus. The partition
function should be compared with that of free type II strings.  In all
three cases, we obtain an identical result.

An essential tool that we use is the thermodynamic partition function.
In our case, we simply regard it as a generating function for the
energy spectrum of free string theory.  Of course, a certain amount of
caution must always be used when discussing string theory at finite
temperature.  Closed string theory is a theory of quantum gravity.  In
a gravitational system, a homogeneous state with finite energy density
has the Jeans instability where it collapses to form black holes.  The
Jeans instability sets in at long wave-lengths so one might expect
that the volume of a system in which conventional thermodynamics is
valid is limited so as to cut off those wave-lengths.  In fact, the
limiting volume is greater if the string coupling is less.  We can get
a crude estimate of the bound by comparing the entropy of black hole
states with that of a thermal gas of gravitons.

Black holes have a large amount of entropy and can easily dominate the
density of states of a quantum gravitational system.  A perfect gas of
gravitons has energy and entropy given roughly by
$$E=T^DV~~$$
and 
$$~~S=T^{D-1}V$$ respectively.  A black hole with energy $E$ has
Schwarzchild radius $R\sim E^{1/(D-3)}$ (in Planck units) and entropy
proportional to the area of its event horizon, $S\sim
(E)^{\frac{D-2}{D-3}}$.  In the set of all possible states, the
perfect gas states dominate over black hole states if they have more
entropy,
$$
T^{D-1}V> (E)^{\frac{D-2}{D-3}}\approx ( T^D V )^{\frac{D-2}{D-3}}
$$
which,  with linear dimension given by $L=V^{1/(D-1)}$, yields the formula
$$
L< \left(\frac{L_S}{L_P}\right)^{\frac{D-2}{D-1}}
\frac{1}{T^{\frac{2D-3}{D-1}} }
$$
where we have re-scaled the temperature and length so that they are in
string units, rather than Planck units.  Thus, stringy states are
excited if $T\sim 1$ (in fact, to the accuracy of our estimates, the
Hagedorn transition happens when $T\sim 1$).  If the string coupling,
$g_s$ is small, then $L_S/L_P$ is greater than one.  The number of
strings lengths in the maximal volume is roughly $L_S/L_P$ which is
large at weak string coupling.

\section{DLCQ of the string: Path integral}

In this Section we will review a result about discrete light cone
quantization of strings using the path integral which were first
obtained in ref.\cite{Grignani:2000zm}.  The vacuum energy of the
string is computed by the Polyakov path integral\footnote{Here, for
concreteness, we use the Neveu-Schwarz-Ramond superstring.  It is
important to keep in mind that our considerations which lead to the
result of this Section would apply equally well to the bosonic sector
of any string theory, including the bosonic string.  The relationship
with matrix theory which we shall discuss later, the most appropriate
theory is the Green-Schwarz superstring.}
\begin{equation}
F=-
\sum_{g,\sigma}g_s^{2g-2} \int [dh_gdXd\Psi]~e^{-S[X,\Psi,h]}
\label{partf}
\end{equation}
where the action is 
$$
S=
\frac{1}{4\pi\alpha'}\int \sqrt{h}\left(\partial_aX^\mu\partial^a
X^\mu-2\pi i\alpha'\Psi^\mu\gamma\cdot\nabla\Psi^\mu\right)
$$
The string coupling constant is $g_s$ and its powers weight the genus,
$g=0,1,\ldots$, of the string's worldsheet.  There is also a sum over
spin structures, $\sigma$ which, with the appropriate weights, imposes
the GSO projection.  For each value of the genus, $g$, $[dh_g]$ is an
integration measure over all metrics of that genus and is normalized
by dividing by the volume of the worldsheet re-parameterization and
Weyl groups.

To get finite temperature $T=1/\beta$ we compactify the Euclidean time,
\begin{equation}
\left( X^0,\vec X, X^9\right)\sim \left(X^0+\beta, \vec X, X^9\right)
\label{comp2}
\end{equation}
This also requires appropriate modification of the GSO projection to
make the target space fermions anti-periodic.

DLCQ requires compactification of the light-cone in Minkowski
space.  We do this by making the identification of Minkowski space
coordinates
\begin{equation}
\frac{1}{\sqrt{2}}\left(t+x^9\right)\sim
\frac{1}{\sqrt{2}}\left(t+x^9\right)+2\pi R
\end{equation}  
In Euclidean space this is the complex identification
\begin{equation}
\left( X^0,\vec X, X^9\right)\sim \left( X^0, \vec X,
X^9\right)+\sqrt{2}\pi R\left( i,0,1\right)
\label{comp1}
\end{equation}
With this compactification the GSO projection is unmodified.  
  
In order to implement compactifications, we must include topological
sectors in the path integral.  In these sectors, the string worldsheet
wraps the compact directions.  To study the general case, we begin by
developing some notation.  The worldsheet is a Riemann surface
$\Sigma_g$ of genus $g$ whose homology group $H_1(\Sigma_g)$ is
generated by the a complete set of cycles (non-contractable closed
curves) which can always be separated into two sets and chosen to have
the canonical intersection property
\begin{eqnarray}&&
a_1,a_2,\ldots, a_g~,~b_1, b_2, \ldots, b_g  \nonumber\\
&&
a_i\cap a_j=\emptyset~,~b_i\cap
b_j=\emptyset~,~ a_i\cap b_j=\delta_{ij}
\label{homology}
\end{eqnarray}
In addition, there is a basis of holomorphic differentials
$\omega_i\in H^1(\Sigma_g)$ (1-forms obeying $d\omega^i=0$) with the
properties
\begin{equation}
\oint_{a_i}\omega_j=\delta_{ij}
~~~,~~~
\oint_{b_i}\omega_j=\Omega_{ij}
\label{orthog}
\end{equation}
where $\Omega=\Omega_1+i\Omega_2$ is the period matrix.  
It is complex, symmetric and has positive definite imaginary part.

When the worldsheet wraps a compact dimension, the bosonic coordinates
of the string $X^\mu$ should have a multi-valued part which changes by
$\beta\cdot$integer$\cdot(1,0,0)$ or $\sqrt{2}\pi
R\cdot$integer$\cdot(i,0,1)$ as it is moved along a homology cycle. To
deal with single-valued quantities, we consider the 1-forms $dX^\mu$
which can be expanded in holomorphic and anti-holomorphic
differentials and exact parts,
\begin{eqnarray}
dX^0=\sum_{i=1}^{g}\left( \lambda_i\omega_i+\bar\lambda_i\bar\omega_i\right)+{\rm exact}
\nonumber \\
d\vec X={\rm exact}
\nonumber \\
dX^9=\sum_{i=1}^{g}\left( \gamma_i\omega_i+\bar\gamma_i\bar\omega_i\right)+{\rm exact}
\label{coord}
\end{eqnarray}
The holomorphic and anti-holomorphic 1-forms account for the holonomy.
Indeed, we require that
\begin{eqnarray}
\oint_{a_i}dX^0=\beta n_i+\sqrt{2}\pi Ri p_i
\nonumber \\
\oint_{b_i}dX^0=\beta m_i+\sqrt{2}\pi Ri q_i
\end{eqnarray}
\begin{equation}
\oint_{a_i}dX^9=\sqrt{2}\pi R p_i
~~~,~~~
\oint_{b_i}dX^9=\sqrt{2}\pi R q_i
\end{equation}
With eqn.(\ref{orthog}), we use these equations to solve for the constants in
eqn.(\ref{coord}). Then, with the formula
\begin{eqnarray}
\int\omega_i \bar\omega_j=
\sum_{k=1}^g\left(\oint_{a_k}\omega_i\oint_{b_k}\bar\omega_j-
\oint_{b_k}\omega_i\oint_{a_k}\bar\omega_j\right)=
\nonumber \\
=
-2i\left(\Omega_2\right)_{ij}
\end{eqnarray}
we can find the part of the string action which contains the winding
integers,
\begin{eqnarray}
S=\frac{\beta^2}{4\pi\alpha'}\left(
n\Omega^{\dagger}-m\right)\Omega_2^{-1} \left( \Omega n-m\right)+ 
\nonumber \\
+2\pi
i \frac{ \sqrt{2}\beta R} {4\pi\alpha'}\frac{1}{2}\left[\left(
p\Omega^{\dagger}-q\right)\Omega_2^{-1} \left( \Omega n-m\right)
\right. \nonumber \\ \left.  +\left(
n\Omega^{\dagger}-m\right)\Omega_2^{-1} \left( \Omega
p-q\right)\right]+\ldots
\end{eqnarray}
The integers $p_i$ and $q_i$ appear linearly in a purely imaginary
term in the action. Furthermore, since they come from the
compactification of the light cone, this is the only place that they
will appear in the entire string path integral (unlike $n_i$ and $m_i$
which, because of the modification of the GSO projection at finite
temperature, should appear in the weights of the sum over spin
structures).  

The action must be exponentiated and summed over $p_i$ and $q_i$.
There are two cases to consider.  First there is the special case
where all $m_i=0=n_i$.  This gives the zero temperature limit of the
free energy, that is, the vacuum energy times the space-time volume.
In this case, entire topological part of the action vanishes and,
after dividing by the space-time volume, the vacuum energy density
does not depend on the null compactification radius, $R$.  The
summation over $p_i$ and $q_i$ give a divergent overall factor.  

The fact that the vacuum energy density does not depend on $R$ can be
understood from the idea of Seiberg and Sen
\cite{ns}-\cite{Sen:1997we} together with T-duality.  The space-time
with a compact null dimension can be obtained by beginning with a
space-time with a compact spatial dimension, then doing an infinite
boost along that direction and simultaneously shrinking the size of
the original space-like compactification radius to zero.  However,
because of T-duality, in closed string theory, the sum of zero point
energies of the strings at zero compactification radius is just
identical to that sum at infinite compactification radius.  The
partition function computes energies, not energy densities.  Thus, the
vacuum energy density must be unaffected by DLCQ and the infinite
factor which arises is just the ratio of the radii of the infinite and
zero sized circles.  This can easily be seen by explicitly working
through the exercise of using a boosted spatial compactification
$$
X^+\sim X^+ +\sqrt{2}\pi R^1\Lambda
~,~
X^-\sim X^--\sqrt{2}\pi R^1/\Lambda
$$
where $\Lambda$ is the boost parameter and seeing that in the limit of
infinite boost $\Lambda\to\infty$ and vanishing spatial radius $R^1\to
0$ such that the product $R^1\Lambda=\sqrt{2}R$ is held constant, the
vacuum energy divided by $R_1$ (and other $R^1$-independent factors)
to get energy density becomes $R$ independent.

In the following we will consider the situation where the zero
temperature limit of the free energy, the vacuum energy, has been
subtracted.  Then we can assume that at least some of the integers
$n_i,m_i$ are non-zero.  Then, the result of summing the exponential
of the action over $p_i,q_i$ will be periodic Dirac delta functions.
It can be shown that these delta functions impose a linear constraint
on the period matrix of the worldsheet.  The net effect is to insert
into the path integral measure the following expression,
\begin{eqnarray}
\sum_{mnrs}e^{-\frac{\beta^2}{4\pi\alpha'}\left(
n\Omega^{\dagger}-m\right)\Omega_2^{-1} \left( \Omega n-m\right)}
\nu^{-2g}\left| \det\Omega_2\right| 
\cdot \nonumber \\ \cdot\prod_{j=1}^g
\delta\left(\sum_{i=1}^g\left( n_i+\frac{i}{\nu} r_i\right)\Omega_{ij}-\left(
m_j+\frac{i}{\nu} s_j\right) \right)\cdot
\nonumber \\
\cdot U(m,n,\sigma)
\label{mod}
\end{eqnarray}
where $ \nu=\sqrt{2}\beta R/4\pi\alpha'$ is a fixed constant and
$U(m,n,\sigma)$ makes the appropriate modification of the GSO
projection. This modification is discussed with explicit examples for
genus 1 in ref. \cite{Atick:1988si}, for example.  The only other
place that any information about the compactification enters the
theory is in the integration over the zero modes in the bosonic part
of the path integral.  This produces a factor of the total volume of
the space-time which should be divided out of both sides of
eqn.(\ref{partf}) to obtain the thermodynamic free energy density.
  
{\it This is our central result, that the net effect of
compactifications is to simply insert the expression eqn.(\ref{mod})
into the path integral measure.}  This result was first obtained in
ref.\cite{Grignani:2000zm}.  Since the first homology group of the
2-sphere is trivial, the first order in perturbation theory which is
affected is at genus 1.  A similar constraint on the modular
parameters at genus 1 was found for scattering amplitudes and argued
to also exist and resemble eqn.(\ref{mod}) at higher genera in
ref.\cite{Bilal:1998vq}-\cite{Bilal:1998ys}.  It would be interesting
to generalize our further observations about the geometric
interpretation of this constraint (in Section 5) to scattering
amplitudes.

\section{Partition function of the Bosonic string at genus 1}

It is interesting to see what the result of the previous Section tells
us about the DLCQ finite temperature partition function of the Bosonic
string at genus 1.  The usual torus amplitude of the bosonic string is
$$
\frac{F}{V}=-\int_{\cal F}\frac{d\tau_1d\tau_2}{\tau_2}\left( \frac{1}
{4\pi\alpha'\tau_2}\right)^{13}\vert\eta(\tau)\vert^{-48}
$$
where the complex modular parameter $\tau=\tau_1+i\tau_2$ is integrated over
the fundamental domain of the torus,
\begin{equation}{\cal F}~\equiv~\left\{ \tau\left|
|\tau_1|\leq\frac{1}{2};~
\vert\tau\vert\geq 1; \tau_2>0 \right. \right\}
\label{fund}
\end{equation}
Our observations in the previous Section tell us that we obtain the
finite temperature DLCQ partition function by inserting the expression
in eqn.(\ref{mod}) with $g=1$ into the integrand (and set U=1).  When
$g=1$ that expression takes the form
$$
\sum_{mnrs}\frac{\tau_2}{\nu^2} e^{-\frac{\beta^2|\tau m-n|^2}{4\pi\alpha'\tau_2}}
\delta\left(\left(n+\frac{i}{\nu} r\right)\tau-\left((m+\frac{i}{\nu} s\right)\right)
$$
The insertion results in
\begin{equation}
\frac{F}{V}=-\sum_{\tau\in{\cal F}}
\frac{ \nu^{-2} e^{-\frac{\beta^2|n\tau-m|^2}{4\pi\alpha'\tau_2}} } {n^2+\nu^{-2}r^2}
\left(\frac{1}{4\pi\alpha'\tau_2}\right)^{13}\vert\eta(\tau)\vert^{-48}
\label{freenergy5}
\end{equation}
where
\begin{equation}\tau=\frac{m+\frac{i}{\nu} s}{n+\frac{i}{\nu} r}
\label{tau4}
\end{equation}
In eqn.(\ref{freenergy5}) we have integrated the delta function (with the
necessary Jacobian).  Here, we see that the moduli space of the torus
is completely discretized.  We will find a geometric interpretation of
this discretization in Section 5.  In the next Section we pause
to see whether this result is consistent with what one would obtain by
an alternative technique - working out the spectrum of the free DLCQ
string by operator methods and constructing the thermodynamic
partition function by tracing the appropriate density matrix.

\section{Operator derivation of the partition function for
bosonic string at genus 1}

In this Section, in order to check the validity of eqn.(\ref{freenergy5}),
we will compute the genus 1 contribution to the finite temperature
partition function using operator quantization in the (discrete)
light-cone gauge.  For simplicity, we will use the bosonic string as
an example.  This derivation as well as one for the type II
superstring which will be discussed in the following Section was given
in more detail in \cite{Grignani:1999sp}.

In light-cone quantization the mass spectrum of the Bosonic string on
26-dimensional Minkowski space is given by the eigenvalues of the mass
operator
\begin{equation}
M^2=2P^+P^- -\vec P^2=\frac{2}{\alpha'}\left(L_0+\tilde L_0-2\right)
\label{spectrum}
\end{equation}
which also satisfy the level matching condition
$$
L_0-\tilde L_0=0$$
where
$$
L_0=\frac{1}{2}\sum_{n-1}^\infty\alpha_{-n}\cdot\alpha_n
~~~,~
\tilde L_0 = \frac{1}{2}\sum_{n=1}^\infty\tilde\alpha_{-n}
\cdot\tilde\alpha_n
$$
and 
$$
\left[ \alpha^i_m, \alpha^j_n\right]=m\delta_{m+n,0}
\delta^{ij}
~,~
\left[\tilde \alpha^i_m, \tilde\alpha^j_n\right]=m\delta_{m+n,0}
\delta^{ij}
$$
with $i,j=1,...,24$ are the usual string oscillators.    
The light-cone momenta are
$$
P^\pm=\frac{1}{\sqrt{2}}\left(P^0\pm P^{D-1}\right)
~~~,~
\vec P=(P^1,\ldots,P^{D-2})
$$
When we compactify the null direction to get DLCQ of the string we
must make the coordinate identification
$$
X^+\sim X^+ + 2\pi R
$$
This modifies the spectrum of the string in two ways.  First, it quantizes
the momentum conjugate to $X^+$, (since $[X^{+},P^{-}]=i$), 
$$
P^-=N/R
$$
Second, it introduces a wrapping number $w$ of the string world sheet
on the compact dimension.  It turns out that this wrapping number does
not appear in the mass shell condition, eqn.(\ref{spectrum}), which
remains unmodified.  The only modification is to the level matching
condition which becomes 
\begin{equation}
L_0-\tilde L_0=Nw
\label{matching}
\end{equation}
The usual light-cone quantization is recovered if we let $R\to \infty$ and
$N\to\infty$ holding $P^-$ fixed.  Then the spectrum of $P^-$ reverts to a
continuum and states with $w\neq0$ go to infinite mass.

Here we will consider the thermodynamic partition function of the free
string.  We will begin by deriving it using the spectrum which we
discussed above.  The spectrum of the free string is identical to the
spectrum of a tower of free particles.  The free energy of a second
quantized free bosonic particle of mass $M$ is given by
\begin{equation}
F= \frac{1}{\beta}{\rm Tr}\ln\left( 1-e^{-\beta P^0}\right)
\label{freenergy2}
\end{equation}
where we have subtracted the temperature independent vacuum energy.
Our first step is to Taylor expand this expression to produce
\begin{equation}
F=-\sum_{k=1}^\infty \frac{1}{k\beta} {\rm Tr} e^{-k\beta P^0}
\end{equation}
Here, we see that the many particle free energy is expanded in terms
of heat kernels which propagate some positive number of intervals 
$\beta$ in imaginary time.  

The energy in terms of light-cone momenta is 
\begin{equation}
P^0=\frac{\left(P^++P^-\right)}{\sqrt{2}}=
\frac{1}{\sqrt{2}}\left( \frac{N}{R}+\frac{ \vec P^2+M^2}{2N/R}\right)
\end{equation} 
We see from this expression that, if the energy is to be positive,
then $N$ must also be positive.  Furthermore, there is no finite
energy physical state with $N=0$.
 
The free energy density is 
\begin{equation}
\frac{ F}{V}
=-\int \frac{d\vec p}{(2\pi)^{\small D-2} }  
\sum_{k,N=1}^\infty
\frac{ e^{ -\frac{\beta kN}{\sqrt{2}R} 
-\frac{ \vec p^2+M^2}{2\sqrt{2}N/k\beta R} } }{\sqrt{2}\pi \beta Rk}
\label{freenergy3}
\end{equation} 
We have divided both sides by the spatial volume which is
$V$=$\sqrt{2}\pi R$(transverse volume).  We have taken a trace over
the momenta.  The spectrum of $P^+$ is discrete and tracing over it
leads to the sum over $N$.  The integral over transverse momenta $\vec
p$ with measure (transverse volume)$\cdot\int {d^{\small D-2}p
}{(2\pi)^{\small D-2}}$ is the trace over the spectrum of $\vec P$.
Finally, integrating over $\vec p$, we get
\begin{equation}
\frac{F}{V}=-\sum_{k,N=1}^\infty \left(
\frac{N}{\sqrt{2}\pi k\beta R}\right)^{D/2}
\frac{e^{-\frac{\beta Nk}{\sqrt{2} R}-\frac{\beta R k}{2\sqrt{2} N}M^2}
}{N}
\label{pf32}
\end{equation}

We can derive the string partition function from this expression by
summing it over the masses of physical particle states which occur in
the free string spectrum.  In order to do this, we use the mass shell
condition, eqn.(\ref{spectrum}), to write the mass-dependent part
of eqn.(\ref{freenergy3}) as an operator and take its trace over physical
states of the string.  This means evaluating the trace over states
which obey the level matching condition eqn.(\ref{matching}) of the
operator
$$ \exp(-2\pi \tau_2(L_0+\tilde L_0-2))$$ where $\tau_2=
\frac{k}{N}\frac{\sqrt{2}\beta R}{4\pi\alpha'}=\frac{k}{N}\nu$ with
$\nu$ the quantity which was defined after eqn.eqn.(\ref{mod}). 
  
We project onto that part of the spectrum which
obeys the level matching condition using a fourier integral,
$$
\delta\left(L_0-\tilde L_0-wN\right)
=\int_{0}^{1}d\tau_1 e^{2\pi i\tau_1(L_0-\tilde L_0-Nw)}
$$
so that the sum over physical masses is equivalent to 
computing 
\begin{equation}
\int_0^1 d \tau_1  \vert {\rm Tr} e^{2\pi i\tau {L_0} } \vert^2
e^{-2\pi i\tau_1 wN+4\pi\tau_2}
\end{equation}
This expression is the only place where the wrapping number $w$ will
appear and it should be summed over all integers.  When summed, it
gives a periodic delta function,
$$
\sum_{w=-\infty}^\infty \exp(- 2\pi i \tau_1 wN)
=\frac{1}{N} \sum_{r=0}^{N-1} \delta(\tau_1-r/N)
$$
In this way, the sum over physical masses takes the form
\begin{equation}
\frac{1}{N}
\sum_{r=0}^{N-1}  \vert {\rm Tr}
\exp\left(2\pi i\tau {L_0}\right) \vert^2_{\tau_1=r/N}
e^{4\pi\tau_2}
\end{equation}
with
\begin{equation}
\tau=\tau_1+i\tau_2= \frac{r+i k \nu}{N}
\label{tau3}
\end{equation}
The trace over the spectrum of $L_0$ is 
given in terms of the Dedekind eta-function
$$
{\rm Tr}e^{2\pi i\tau L_0}
=\prod_n\left(\frac{1}{1-e^{2\pi i \tau n}}\right)^{24}
\equiv e^{2\pi i\tau}\eta^{-24}(\tau)
$$
The final result for the thermodynamic partition function of the
Bosonic string at temperature $1/\beta$ with light-cone radius $R$ is
\begin{eqnarray}
\frac{ F}{V}=  
-\sum_{k,N=1}^\infty\sum_{r=0}^{N-1}\frac{1}{N^2}\left(
\frac{1 }{4\pi^2\alpha'\tau_2}\right)^{13}\cdot
\nonumber\\
\cdot
e^{-\frac{k^2\beta^2}{4\pi\alpha'\tau_2}}\vert \eta(\tau)\vert^{-48}
\label{freenergy4}
\end{eqnarray} 
(We have put D=26.)  This expression is not yet identical to
eqn.(\ref{freenergy5}).  It is summed over three integers which
specify the modular parameter in eqn.(\ref{tau3}) instead of the four
integers which specify the modular parameter in eqn.(\ref{tau4}) and
the two parameters have a different domain.  Indeed, they can be shown
to be equal using the ``Polchinski trick'' \cite{Polchinski:zf}.  The
essential observation is that the summand 
$$
\left(\frac{1}{4\pi\alpha'\tau_2}\right)^{13}\vert\eta(\tau)\vert^{-48}
$$
is invariant under modular transformations
$$
\tau\to \frac{  a+b\tau}{c+d\tau}~~abcd\in{\cal Z}~~ad-bc=1
$$
We can always find a modular
transformation which sets the integer $n$ in eqn.(\ref{tau4}) and
eqn.(\ref{freenergy5}) to zero.  This modular transformation also maps
the point $\tau=\frac{ m+\frac{i}{\nu}s }{ n + \frac{i}{\nu} r }$ in
the fundamental domain to a point $\tau'=\frac{ m'+\frac{i}{\nu}s'
}{\frac{i}{\nu} r' }$ in the strip $\tau_2\in[0,1]~ \tau_1>0$.
The result is that the sum over the three integers $kNr$ in the domain
$\tau_2\in[0,1]~ \tau_1>0$ is equivalent to the sum over four integers
$mnrs$ over the domain ${\cal F}$.  With this observation,
eqn.(\ref{freenergy4}) and eqn.(\ref{freenergy5}) are seen to be
identical.

\section{Type II superstring at genus 1}

As another concrete example, consider the type II superstring torus amplitude 
\begin{eqnarray}
\frac{F}{V}=-\int_{\cal F}\frac{d^2\tau}{4\tau_2}
\frac{1}{\left(4\pi^2\alpha'\tau_2\right)^5}
\frac{1}{ \left| \eta(\tau)\right|^{24} }\cdot
\nonumber \\
\cdot\left[ \theta_2^4\bar\theta_2^4
+\theta_3^4\bar\theta_3^4 + \theta_4^4\bar\theta_4^4
 +\left( \theta_2^4\bar\theta_4^4+\theta_4^4\bar\theta_2^4
\right)
\right.
\nonumber \\
\left.
-\left( \theta_2^4\bar\theta_3^4+\theta_3^4\bar\theta_2^4
\right) 
 - \left( \theta_3^4\bar\theta_4^4 
+ \theta_4^4\bar\theta_3^4\right)\right]
\label{pf1}
\end{eqnarray}
where $\theta_k(0,\tau)$ are Jacobi theta functions and $\eta(\tau)$
is the Dedekind eta-function. 
The modification of this formula by the null compactification can be found using
eqn.(\ref{mod}),
\begin{eqnarray}
\frac{F}{V}=-\sum_{\tau\in{\cal F}} \frac{\nu^{-2}}{n^2+\nu^{-2}r^2} 
\frac{e^{-\frac{\beta^2 \vert n\tau-m\vert^2}{4\pi\alpha'\tau_2}}
}{\left(4\pi^2\alpha'\tau_2\right)^5}
\frac{1}{4\left| \eta(\tau)\right|^{24}}\cdot
\nonumber \\
\left[  \theta_2^4\bar\theta_2^4
+\theta_3^4\bar\theta_3^4 + \theta_4^4\bar\theta_4^4
+(-1)^{m+n}\left( \theta_2^4\bar\theta_4^4+\theta_4^4\bar\theta_2^4
\right)
\right.
\nonumber \\
\left.
-(-1)^n\left( \theta_2^4\bar\theta_3^4+\theta_3^4\bar\theta_2^4
\right) 
- (-1)^m\left( \theta_3^4\bar\theta_4^4 
+ \theta_4^4\bar\theta_3^4\right)\right]
\label{pf} 
\end{eqnarray}
where $
\tau= (m+\frac{i}{\nu} s)/(n+\frac{i}{\nu} r)
$
and one should sum over the integers so that $\tau$ is in the
fundamental domain, ${\cal F}$.  It was shown in \cite{Grignani:1999sp}
that this is precisely the finite temperature partition function that is
obtained for the type II superstring by operator methods and DLCQ.

Part of the argument showed that it is also equal to the partition
function of the Green-Schwarz formulation of the type II superstring.
The equivalence is demonstrated by using modular transformations and
identities for theta functions.  Another presentation of the same
function, which we shall have use for later is to rewrite eqn.(\ref{pf})
as the Hecke operator \cite{serre} acting on the partition function of
a superconformal field theory, with torus worldsheet and target space
$R^8$:
\begin{eqnarray}
\frac{F}{V}&=&-\frac{1}{\sqrt{2}\pi R\beta}{\cal
H}[e^{-\beta/\sqrt{2}R}]*
\nonumber \\
&&\left[\left(\frac{1}{4\pi^2\alpha' \tau_2}\right)^4 
\frac{1}{\left| \eta(\tau)\right|^{24}}\left|
\theta_2(0,\tau)\right|^8\right]_{\tau=i\nu}
\label{pf4}
\end{eqnarray}
The factor in front is the ratio of volumes of $R^8$ and $R^9\times
S^1$ with compactified light cone.  The action of ${\cal H}[p]$ on a
function $\phi(\tau,\bar\tau)$ is defined by
\begin{equation}
{\cal H}[p]*\phi(\tau,\bar\tau)=\sum_{N=0}^\infty \frac{p^N}{N}
\sum_{\stackrel {kr=N}  {s~{\rm mod}~k} }
\phi\left(  \frac{r\tau+s}{k}, \frac{ r\bar\tau +s}{k}\right)
\end{equation}
where the sum is over $k$ and $r$ which are divisors of $N$ with the
restriction that $r$ is an odd integer.  This operation appears in the
quantization of superconformal field theories on symmetric orbifolds
\cite{Dijkgraaf:1996xw}-\cite{Fuji:2000fa}. We will construct an explicit
version of it in our discussion of matrix strings in a later Section.

\subsection{Geometrical interpretation of the 
constraint on Riemann surfaces}

In the sectors with non-zero wrappings, the delta function in
eqn.(\ref{mod}) restricts the integration over metrics in the string path
integral to those for which the period matrix obeys the constraint
\begin{equation}
\sum_{i=1}^g\left(n_i+\frac{i}{\nu} r_i\right)\Omega_{ij}-\left(
m_j+\frac{i}{\nu} s_j\right)=0
\label{const}
\end{equation}
for all combinations of the $4g$ integers $m_i,n_i,r_i,s_i$ such that
$\Omega$ is in a fundamental domain.  This is clearly a subset of all
Riemann surfaces.  It is interesting to ask whether one can
characterize this subset in a convenient way.  In this Section we will
review a theorem about the classification which was proved in
\cite{Grignani:2000zm}.

A hint about this classification can be found by counting the number
of moduli of the Riemann surface that eqn.(\ref{const}) removes.  The
number of complex moduli of a Riemann surface $\Sigma_g$ is $3g-3$
when $g>1$.  Since the columns of the period matrix are linearly
independent vectors, eqn.(\ref{const}) contains $g$ independent complex
constraints.  Thus the complex dimension $3g-3$ is reduced to $2g-3$
and there is further discrete data contained in the integers. 
\footnote{This
could alternatively be viewed as a partial discretization of the
moduli space. One would expect that, when the compactifications are
removed, either $\beta\rightarrow\infty$ or $R\rightarrow\infty$, the
discrete data assembles itself to a ``continuum limit'' which restores
the $g$ complex dimensions of moduli space.}

$2g-3$ is the dimension of the moduli space of a genus $g$ branched
cover of a torus.  Indeed, given $n$ branch points, the genus is one
plus the number of branch cuts, $g=1+n/2$.  The moduli are the
independent positions of the branch points, $n-1$ where one is
subtracted for overall translation invariance.  The resulting
dimension is $2g-3$, identical to the one which we found above.
  
In fact, we can prove the following
\vspace{5pt}
\noindent
{\bf Theorem}: $\Sigma_{g}$ is a branched cover of a 2-torus $T^2$
with modular parameter $\frac{i}{\nu}$ if and only if $\Omega_{ij}$
obeys eqn.(\ref{const}), for some choice of integers $m_{i}, n_{i},
r_{i}$ and $s_{i}$.

\vspace{5pt}
\noindent
{\it Proof}: The generators of the first homology group of $T^2$ are two closed
loops $(\alpha,\beta)$ which span the vector space $H_1(T^2, {\cal C})$. 
The dual vector space, the first cohomology group, $H^1(T^2, {\cal C})$ 
is spanned by the basis of holomorphic and anti-holomorphic
differentials $\gamma$ and $\bar\gamma$.  They can be normalized as,
$$
\oint_\alpha \gamma=1~~,~~\oint_\beta \gamma=\frac{i}{\nu}
$$
$$ \oint_\alpha
\bar\gamma=1~~,~~\oint_\beta\bar\gamma=-\frac{i}{\nu}
$$

The Riemann surface $\Sigma_g$ is a branched cover of $T^2$ if there
exists a continuous holomorphic map $f$, such that
\begin{equation}
\def\mapright#1{\smash{\mathop{\longrightarrow}\limits^{#1}}}
\Sigma_g ~\mapright{\quad f\quad}~ T^2
\label{map}
\end{equation}
The map $f$ takes closed loops on $\Sigma_g$ to closed loops on $T^2$.
In particular, the generators eqn.(\ref{homology}) must map as
\begin{equation}
\def\mapright#1{\smash{\mathop{\longrightarrow}\limits^{#1}}}
(a_i,b_j)~~\mapright{f} ~~(n_i\alpha+r_i\beta,m_j\alpha+s_j\beta)
\label{maph}
\end{equation}
for some integers $m_i,n_i,r_i,s_i$.
This gives a mapping between the vector spaces $H_1(\Sigma_g,{\cal C})$ 
and $H_1(T^2,{\cal C})$.  
The pull-back of this mapping on the dual vector spaces 
\begin{equation}
\def\mapright#1{\smash{\mathop{\longrightarrow}\limits^{#1}}}
H^1(T^2,{\cal C}) ~~\mapright{f^*}  ~~H^1(\Sigma_g,{\cal C})
\end{equation}
is defined by its action on the basis,
\begin{eqnarray}
a_i\circ f^* (\gamma)=\oint_{a_i}f^*(\gamma)
\equiv f(a_i)\circ\gamma=
\nonumber \\
=n_i\oint_\alpha \gamma +r_i\oint_\beta \gamma=n_i +\frac{i}{\nu} r_i  \cr
b_j\circ f^* (\gamma )=\oint_{b_j}f^*(\gamma )
\equiv f(b_j)\circ\gamma =
\nonumber \\
=
m_j\oint_\alpha \gamma +s_j\oint_\beta 
\gamma = m_j+\frac{i}{\nu} s_j
\label{pullback}
\end{eqnarray}

Consider the $g$ particular elements of $H_1(\Sigma_g,{\cal C})$,
which we shall call $c_i$.  They are linear combination of the cycles
which are given by
\begin{equation}
c_j=\sum_{i=1}^g a_i\Omega_{ij}-b_j
\end{equation}
It can be checked from the definition of the period matrix
eqn.(\ref{homology}) (and the fact that it is symmetric) that the
operation of any holomorphic differential, $\eta$, annihilates all of
these combinations of cycles,
\begin{equation}
c_j\circ\eta =
\sum^g_{i=1}\Omega_{ij}\oint_{a_i}\eta -\oint_{b_j}\eta =0
\label{ker}
\end{equation}
The holomorphic nature of the mapping, $f$, guarantees that
$f^*(\gamma)$ is a holomorphic differential on $\Sigma_g$.  The
equation eqn.(\ref{ker}) should therefore also apply to $f^*$. Then it
follows that
\begin{eqnarray}
0=c_j\circ f^*(\gamma)=f(c_j)\circ \gamma=
\nonumber \\
=\sum_{i=1}^g (n_i+\frac{i}{\nu} r_i) \Omega_{ij}-(m_j+\frac{i}{\nu} s_j)
\end{eqnarray}
which is the constraint on the period matrix in eq.(\ref{const}).
Thus, we have proven that, if $\Sigma_g$ is a branched cover, then its
period matrix obeys eqn.(\ref{const}).  We have further found an
interpretation of the integers, they are those which appear in the
projections of the homology generators of $\Sigma_g$ onto the homology
generators of $T^2$ which are induced by the covering map.

To prove the converse, we must now show that if the constraint in
eqn.(\ref{const}) is satisfied, then a covering map, $f$, exists.  We
will demonstrate this by explicit construction.  Consider the line
integral of a linear combination of the holomorphic differentials on
$\Sigma_g$,
\begin{equation}
z(P)=\int_{P_0}^P \sum_{k=1}^g \lambda_k\omega_k
\label{z}
\end{equation}
with $P_0$ a fixed base-point.  We want to view this integral as a mapping
between points $P\in\Sigma_g$ and the numbers $z(P)$ which are points
in the complex plane.  Of course, this mapping depends on the path of integration
which is chosen and is multi-valued.  In fact, the best we can do is to 
adjust the mapping so that $z(P)$ is a function from $\Sigma_g$ to the
quotient of the complex plane by a lattice.  

For this, we wish to choose the coefficients $\lambda_k$ such that
this integral defines a map from points $P\in\Sigma_g$ to the torus
$z(P)\in T^2$ whose holomorphic coordinates are the complex numbers
$z(P)$ with the identification 
$$
z\sim z+ p+\frac{i}{\nu} q
$$ 
where $p$
and $q$ are integers.  The integral depends on the path of
integration.  If the path is changed by a combination of homology
cycles of $\Sigma_g$, say $k_ia_i+l_i b_i$ where $k_i,l_i$ are
integers, the integral on the right-hand side of eqn.(\ref{z}) changes
by
$$
\delta z=\sum_j\lambda_j\left( k_j+\sum_i\Omega_{ji}l_i\right)
$$
$\lambda_k$ must be chosen so that this change is integer multiples of 
the periods of $T^2$.  This can easily be done if $\Omega$ obeys
eqn.(\ref{const}). Then, with the choice 
$$\lambda_i=n_i+\frac{i}{\nu} r_i$$
we see that 
$$
\delta z=\sum_j(n_j+\frac{i}{\nu} r_j)k_j+\sum_i(m_i+\frac{i}{\nu} s_i)l_i=
$$
$$
={\rm integer}+\frac{i}{\nu}\cdot{\rm integer}
$$
and we have constructed an explicit covering map, $z(P)$. {\bf Q.E.D.}

\section{The thermodynamic partition function of matrix string theory}

In this Section, we will review the argument of
ref.\cite{Grignani:1999sp} which computes the thermodynamic partition
function of the free string limit of matrix string theory.

Matrix string theory is a 1+1-dimensional maximally supersymmetric
Yang-Mills theory on a spatial circle.  Its Hamiltonian is
(The original derivation is in ref.\cite{Dijkgraaf:1997vv}. For details, see the
reviews in ref.\cite{Taylor:2001vb}\cite{Taylor:1999qk} the arguments in
ref.\cite{Grignani:1999sp}.)
\begin{eqnarray}
H=\frac{R}{2}\int_0^1d\sigma_1{\rm Tr}\left[\Pi_i^2+\frac{1}{4\pi^2
\alpha'^2}(D_{1} X^i)^2 -
\right. \nonumber \\  \left.
-\frac{1}{4\pi^2 \alpha'^3 g_s^2}
\left[X^i,X^j\right]^2
+\frac{E^2}{g_s^2\alpha'} \right.\cr\left. -  
\frac{1}{2\pi \alpha'^{3/2} g_s}\psi^T\gamma_i\left[X^i,\psi\right]-
\frac{i }{2\pi\alpha'}\psi^T \gamma\cdot D_{1}\psi\right]
\label{HDVV}
\end{eqnarray}
Here, $\Pi^i$ is the canonical momentum conjugate to $X^i$, $E$ is the
electric field and canonical conjugate to the gauge field $A_\mu$
which appears in the covariant derivative
$(\partial-i[A,)$ the gauge group is SU(N)
and all degrees of freedom transform in the adjoint representation.
Here, the coordinates have been rescaled so that $\sigma_1\in[0,1)$
and all variables are periodic under the spatial translation,
$X^i(\tau,\sigma_1)=X^i(\tau, \sigma_1+1)$.  This Hamiltonian should be augmented
by a Gauss' law constraint which requires that all physical states are gauge
invariant.

This theory is conjectured to describe M-theory with two
compactification, a spatial dimension with radius $g_s\sqrt{\alpha'}$
which, when the radius is small gives type IIA superstring theory and
a null dimension with radius R to give DLCQ \cite{bfss,s,Dijkgraaf:1997vv}. We
should identify the light-cone momenta of matrix string theory as
$P^-=N/R$ and $P^+=H$ and the quantum states are the gauge invariant
eigenstates of $H$.  The thermodynamic partition function is
\begin{eqnarray}
Z=e^{-\beta F}={\rm Tr} e^{-\beta P^0}
={\rm Tr} e^{-\beta\left(P^++P^-\right)/\sqrt{2}}
\nonumber \\
=
\sum^\infty_{N=0}~e^{-N\beta/\sqrt{2}R^+}~{\rm Tr}\left\{e^{-\beta H/\sqrt{2}}
\right\}
\label{Z}
\end{eqnarray}
The trace of $\exp(-\beta H/\sqrt{2})$ over gauge invariant states has
the standard path integral expression of a finite temperature Yang-Mills theory
partition function 
\begin{equation}
Z[\beta]=\sum_{N=0}^\infty e^{-\frac{\beta N}{\sqrt{2}R}}
\int [dAdXd\psi]~e^{-S_E[A,X^i,\psi]}
\label{pf33}
\end{equation}
where $S_E$ is the Euclidean action. The time variable in the Yang-Mills theory is
Euclidean and compact with domain $\sigma_2\in[0,\beta]$.  
By re-scaling the Euclidean coordinates
so that they live in the domain $\sigma_a\in[0,1)$ we can write the action
as
\begin{eqnarray}
S_E=\frac{1}{4\pi\alpha'}
\int d^2\sigma\sqrt{|g|}{\rm Tr}\left\{ 
g^{ab}D_a X^i D_b X^i+\right.\cr\left. +
\frac{g_s^2\alpha'}{|g|}
F_{12}^2-
\frac{1}{\alpha'g_s^2}
\sum_{i<j}\left[ X^i,X^j\right]^2 - \right.\cr\left. 
-i\psi^T\gamma\cdot D\psi
- \frac{1}{\sqrt{\alpha'}g_s}
\psi^T\gamma^i\left[X^i,\psi\right]
\right\}
\label{eucact}
\end{eqnarray}
We have re-scaled Euclidean time $\sigma_2$ so that the integration is
over a box of area one, $0\leq\sigma_a<1$, and $\beta$ appears as a
factor in various coupling constants which have been absorbed into the
metric. $(\gamma^\mu,\gamma^i)$ are the ten-dimensional Gamma matrices
in the Majorana-Weyl representation.  Bosonic variables are periodic
in both space and time and fermionic variables are periodic in space
and anti-periodic in time.  The anti-periodicity of the fermion field
in the Euclidean time comes from taking the trace in eqn.(\ref{Z}).
Thus, the space that the Yang-Mills theory is defined on is a torus.
The metric of the torus is \cite{Grignani:1999sp}:
\begin{equation}
g_{ab}=\left( \matrix{ 1 & 0\cr 0& \nu^2}\right)
\label{torusmetric}
\end{equation}
where $\nu=\frac{\sqrt{2}\beta R}{4\pi\alpha'}$ is a familiar
parameter.  This is just the metric of the underlying torus which we
found in the previous Sections.  Furthermore, the Yang-Mills coupling
constant is given by $g^2_{YM}= 1/\alpha'g_s$.

The limit $g_s\rightarrow 0$ of this theory was developed by
Dijkgraaf, Verlinde and Verlinde \cite{Dijkgraaf:1997vv}.  The
hypothesis is that the degrees of freedom which become perturbative
strings in the limit $g_s\to 0$ are the simultaneous eigenvalues of
the matrices which must commute if they are to have finite energy.
Commuting matrices can be simultaneously diagonalized by a gauge
transformation
\begin{eqnarray}
X^i(\sigma)&=&U(\sigma)X^i_D(\sigma)U^{-1}(\sigma) \cr
\psi(\sigma)&=&U(\sigma)\psi_D(\sigma)U^{-1}(\sigma) \cr
A_\mu(\sigma)&=&
iU(\sigma)\left(\partial_\mu-iA^D_\mu(\sigma)\right)U^{-1}(\sigma)
\label{diag}
\end{eqnarray}
where $X^i_D$, $\psi_D$ and $A^D_\mu$ are diagonal.  The fields $X^i$,
$\psi$ and $A_\mu$ live on the torus with metric given in
eqn.(\ref{torusmetric}).  Eigenvalues of these matrices solve
polynomial equations with doubly periodic coefficients.  They are
generally not single-valued functions on the torus.  Instead, they are
single-valued functions on branched covers of the torus.  Since they
are all diagonalized by the same unitary transformation, they must all
live on the same branched cover.

In this way we see that the degrees of freedom which correspond to
perturbative strings live on branched covers of a basic torus,
precisely those branched covers which, by the arguments of Section 2
and 5, are the worldsheets of DLCQ strings.  It would be very
interesting to show that an asymptotic expansion of the matrix string
free energy in eqn.(\ref{Z}) reproduces the string perturbation theory
in eqn.(\ref{partf}) with the constraint eqn.(\ref{mod}) inserted.
Although there is a large amount of literature in this direction
\cite{Wynter:1997yb}-\cite{Schiappa:2000dv}, this has yet to be
demonstrated in detail for genera greater than one.  At genus 1, it is
indeed the case, as was shown in ref.\cite{Grignani:1999sp}.  We will
review the essential features of the argument here.

To get genus 1, we simply take the diagonal matrices to live on a
genus 1 (unbranched) cover of the torus.  The unbranched cover arises
when we assume that the diagonal matrices in eqn.(\ref{diag}) are
smooth functions, without branch points.  Then they have boundary
conditions which are periodic\footnote{Keep in mind that Fermions are
anti-periodic in the Euclidean time direction.} up to permutations.
For example, for the scalar fields
\begin{eqnarray}
(X^i_D)_\alpha(\sigma_1+1,\sigma_2)&=
&(X^i_D)_{P(\alpha)}(\sigma_1,\sigma_2) \cr
(X^i_D)_\alpha(\sigma_1,\sigma_2+1)&=
&(X^i_D)_{Q(\alpha)}(\sigma_1,\sigma_2) 
\label{diagbc}
\end{eqnarray}
where $\alpha=1,...,N$, $P(\alpha)$ and $Q(\alpha)$ are permutations.
The same permutation acts on all of the fields since they are all
diagonalized by the same unitary matrix in eqn.(\ref{diag}).

Consistency requires that the two permutations commute,
\begin{equation}
PQ=QP
\end{equation}
The partition function is now given by
\begin{equation} 
Z=\sum^\infty_{N=0}\frac{1}{N!}
e^{-N\beta/\sqrt{2}R}\sum_{\stackrel{P,Q} {PQ=QP} } Z(P,Q)\ ,
\label{Zpq}
\end{equation}  
where
\begin{equation}
Z(P,Q)=\int[dAdXd\psi]e^{-S_{\rm diag}}
\end{equation}
with the boundary conditions (\ref{diagbc}) determined by $P$ and $Q$.
In eqn.(\ref{Zpq}), for each $N$, we have divided by the volume of the
Weyl group, $N!$, which reflects the fact that the eigenvalues are
defined up to a global permutation\footnote{ From the M-theory point
of view, this factor gives Boltzmann statistics to the D0-branes
\cite{Ambjorn:1998zt}.  The possibility of introducing different
weights in the sum over pairs of commuting permutations has been
discussed in~\cite{Billo:1998fb}.  We shall show in what follows that
the correct partition function for the type IIA string emerges from
eqn.(\ref{Zpq}) only when all of the pairs $(P,Q)$ have the same
weight.  It turns out that, at genus 1, the gauge field integral in
eqn.(\ref{Zpq}) is trivial and we will drop it from now on
\cite{Grignani:1999sp}.  There is a beautiful suggestion in
refs.\cite{bbn,bbnt} that at higher genera it gives the string
coupling constant correct factor $g_s^{2g-2}$.}.  The action is
\begin{eqnarray}
S_{\rm diag}=\frac{1}{4\pi\alpha'}
\int d^2\sigma\sqrt{|g|}\sum_{\alpha a=1}^N
\left\{
g^{ab}\partial_a X^i_\alpha \partial_b X^i_\alpha
+ \right.\cr\left. +
\frac{g_s^2\alpha'}{|g|}
\left(\partial_1A_{2\alpha}-\partial_2 A_{1\alpha}\right)^2
-i
\psi^T_\alpha\gamma\cdot\nabla\psi_\alpha
\right\}
\end{eqnarray}
and with the boundary conditions (\ref{diagbc}). 

It is useful at this point to review some of the salient points about
commuting permutations, for which we follow the appendix of
ref.\cite{Billo:1998fb}.  Another useful resource is ref.\cite{Kostov}.

Consider a fixed permutation $P$ of a set of $N$ elements.  It can be
decomposed into cycles which are subsets of the $N$ elements such that
the permutation interchanges elements cyclically inside each
subset. For example, for $N=9$, the permutation which can be denoted
by
$$
\alpha=\left(\matrix{1&2&3&4&5&6&7&8&9\cr 6&4&1&2&5&3&8&9&7\cr}\right)
$$
can be decomposed into the product of disjoint cycles
$$
\alpha=(163)(24)(5)(789)
$$
Consider such a decomposition of $P$.  Suppose that the number of
cycles of length $k$ is $r_k$, so that $N=\sum_k kr_k$.  Clearly, the
minimum value of $k$ is one and the maximum is $N$. $r_k$ are
non-negative integers. (In the above example, $r_1=1,r_2=1,r_3=2$.)
Denote the elements of the set $\{1,2,...,N\}$ by the three index notation
based on how they transform under $P$:
\begin{equation}
a^{k,\gamma}_n  ~~(k=1,...,N);(\gamma=1,...,r_k):(n=1,...,k)
\end{equation}
This element of $\{1,...,N\}$ is the nth element of the $\gamma$'th
cycle of length $k$.  The action of the permutation on this element is
then
\begin{equation}
P\left(a^{k,\gamma}_n\right)=a^{k,\gamma}_{n+1~{\rm mod}~k}
\label{permact}
\end{equation}

Let $Q$ be a permutation which commutes with $P$.  Consider its action on
the cycle $$\left(a_1^{k,\gamma},...,a_k^{k,\gamma}\right)$$ of $P$. 
Eq. (\ref{permact}) implies that
\begin{equation}
Q\left(P\left(a^{k,\gamma}_n\right)\right)
=Q\left(a^{k,\gamma}_{n+1~{\rm mod}~k}\right)
\end{equation}
Since $PQ=QP$, this equation can be written as
\begin{equation}
P\left(Q\left(a^{k,\gamma}_n\right)\right)
=Q\left(a^{k,\gamma}_{n+1~{\rm mod}\ k}\right)
\end{equation}
This means that
$$\left(Q\left( a_1^{k,\gamma}\right), ,...,
Q\left(a_k^{k,\gamma}\right)\right)$$
is a cycle of length $k$ in $P$, i.e. that
$$\left(Q\left( a_1^{k,\gamma}\right), ,... ,
Q\left(a_k^{k,\gamma}\right)\right)=
$$
$$
= \left(a_{s(k,\gamma)}^{k,\pi_k(\gamma)},...,
a_{s(k,\gamma)+k-1~{\rm mod}~k}^{k,\pi_k(\gamma)}\right)$$ 
Hence, there exists a
permutation $\pi_k(\gamma)$ of the $r_k$ elements of the set of cycles
of length $k$ which is induced by $Q$.  Also, besides permuting the
cycles of $P$ of equal length, $Q$ can also do a cyclic permutation of
each cycle by the integer $s(k,\gamma)$ which can take the values $1,..,k$.   
Thus, the only allowed action of $Q$ is 
\begin{equation}
Q\left( a^{k,\gamma}_n\right)=
a^{k,\pi_k(\gamma)}_{n+s(k,\gamma){~\rm mod}~k}
\label{defnq}
\end{equation}

Consider a set of diagonal components of fields 
obeying the boundary conditions
\begin{eqnarray}
X^i_{\alpha}(\sigma_1+1,\sigma_2)=X^i_{P(\alpha)}(\sigma_1,\sigma_2)\ ,\cr
X^i_{\alpha}(\sigma_1,\sigma_2+1)= X^i_{Q(\alpha)}(\sigma_1,\sigma_2)\ .
\label{pq}
\end{eqnarray}
Consider those  which occur in the cycles
of length $k$ of $P$ and re-label them according to elements
of the cycle,
\begin{eqnarray}
X^i_{a^{k\gamma}_n}(\sigma_1+1,\sigma_2)=X^i_{a^{k\gamma}_{n+1}}
(\sigma_1,\sigma_2)\ ,\cr
X^i_{a^{k\gamma_n}}(\sigma_1,\sigma_2+1)=
X^i_{a^{k\pi(\gamma)}_{n+s}}(\sigma_1,\sigma_2)\ .
\label{pq1}
\end{eqnarray}
For each fixed $\gamma$, we fuse these fields together into
a single function which has the property
\begin{eqnarray}
X^i_{a^{k\gamma}}(\sigma_1+k,\sigma_2)=X^i_{a^{k\gamma}}
(\sigma_1,\sigma_2)\ ,\cr
X^i_{a^{k\gamma}}(\sigma_1,\sigma_2+1)=
X^i_{a^{k\pi(\gamma)}}(\sigma_1+s(k,\gamma),\sigma_2)\ .
\label{pq2}
\end{eqnarray}
Then, we consider a cycle of the permutation $Q$, which must be a subset
of the $r_k$ k-cycles of $P$.  Say this cycle is of length $r$ where 
$1\leq r\leq r_k$.  Then we fuse $r$ of the above fields together to 
get the single field which has completely periodic boundary conditions
\begin{eqnarray}
X^i_{a^{k,r}}(\sigma_1+k,\sigma_2)=X^i_{a^{k,r}}
(\sigma_1,\sigma_2)\ ,\cr
X^i_{a^{k,r}}(\sigma_1,\sigma_2+r)=
X^i_{a^{k,r}}(\sigma_1+s,\sigma_2)\ .
\label{pq3}
\end{eqnarray}  
where $s=\sum_\gamma s(k,\gamma)~{\rm mod}~k$ is the accumulated shift
for the $r$ elements in the cycle of $Q$.

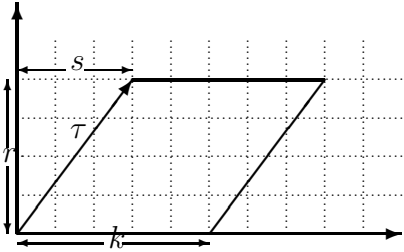
\begin{figure}[htb]
\begin{center}
\setlength{\unitlength}{0.00033300in}%
\begingroup\makeatletter\ifx\SetFigFont\undefined%
\gdef\SetFigFont#1#2#3#4#5{%
  \reset@font\fontsize{#1}{#2pt}%
  \fontfamily{#3}\fontseries{#4}\fontshape{#5}%
  \selectfont}%
\fi\endgroup%
\begin{picture}(6258,3858)(3376,-4186)
\thinlines
\put(3601,-2161){\makebox(1.8519,12.9630){\SetFigFont{5}{6}{\rmdefault}{\mddefault}{\updefault}.}}
\put(3601,-1636){\makebox(1.8519,12.9630){\SetFigFont{5}{6}{\rmdefault}{\mddefault}{\updefault}.}}
\put(4651,-1411){\vector( 1, 0){750}}
\put(4351,-1411){\vector(-1, 0){750}}
\put(5251,-4111){\makebox(1.8519,12.9630){\SetFigFont{5}{6}{\rmdefault}{\mddefault}{\updefault}.}}
\put(3451,-2611){\vector( 0, 1){1050}}
\put(3451,-2911){\vector( 0,-1){1050}}
\put(4951,-4111){\vector(-1, 0){1350}}
\put(5251,-4111){\vector( 1, 0){1350}}
\thicklines
\put(3601,-3961){\vector( 0, 1){3600}}
\put(3601,-3961){\vector( 1, 0){6000}}
\put(5401,-1561){\line( 1, 0){3000}}
\thinlines
\multiput(3601,-3361)(89.55224,0.00000){68}{\makebox(1.8519,12.9630){\SetFigFont{5}{6}{\rmdefault}{\mddefault}{\updefault}.}}
\multiput(4801,-961)(0.00000,-90.90909){34}{\makebox(1.8519,12.9630){\SetFigFont{5}{6}{\rmdefault}{\mddefault}{\updefault}.}}
\multiput(5401,-961)(0.00000,-90.90909){34}{\makebox(1.8519,12.9630){\SetFigFont{5}{6}{\rmdefault}{\mddefault}{\updefault}.}}
\put(3676,-2761){\makebox(1.8519,12.9630){\SetFigFont{5}{6}{\rmdefault}{\mddefault}{\updefault}.}}
\multiput(6001,-961)(0.00000,-90.90909){34}{\makebox(1.8519,12.9630){\SetFigFont{5}{6}{\rmdefault}{\mddefault}{\updefault}.}}
\put(4426,-2461){\makebox(0,0)[lb]{\smash{\SetFigFont{12}{14.4}{\rmdefault}{\mddefault}{\updefault}$\tau$}}}
\multiput(6601,-961)(0.00000,-90.90909){34}{\makebox(1.8519,12.9630){\SetFigFont{5}{6}{\rmdefault}{\mddefault}{\updefault}.}}
\multiput(7201,-961)(0.00000,-90.90909){34}{\makebox(1.8519,12.9630){\SetFigFont{5}{6}{\rmdefault}{\mddefault}{\updefault}.}}
\multiput(7801,-961)(0.00000,-90.90909){34}{\makebox(1.8519,12.9630){\SetFigFont{5}{6}{\rmdefault}{\mddefault}{\updefault}.}}
\multiput(8401,-961)(0.00000,-90.90909){34}{\makebox(1.8519,12.9630){\SetFigFont{5}{6}{\rmdefault}{\mddefault}{\updefault}.}}
\multiput(9001,-3961)(0.00000,90.90909){34}{\makebox(1.8519,12.9630){\SetFigFont{5}{6}{\rmdefault}{\mddefault}{\updefault}.}}
\multiput(3601,-2761)(89.55224,0.00000){68}{\makebox(1.8519,12.9630){\SetFigFont{5}{6}{\rmdefault}{\mddefault}{\updefault}.}}
\multiput(3601,-2161)(89.55224,0.00000){68}{\makebox(1.8519,12.9630){\SetFigFont{5}{6}{\rmdefault}{\mddefault}{\updefault}.}}
\multiput(3601,-1561)(89.55224,0.00000){68}{\makebox(1.8519,12.9630){\SetFigFont{5}{6}{\rmdefault}{\mddefault}{\updefault}.}}
\thicklines
\put(6601,-3961){\line( 3, 4){1800}}
\put(3601,-3961){\vector( 3, 4){1800}}
\thinlines
\multiput(4201,-961)(0.00000,-90.90909){34}{\makebox(1.8519,12.9630){\SetFigFont{5}{6}{\rmdefault}{\mddefault}{\updefault}.}}
\put(5026,-4186){\makebox(0,0)[lb]{\smash{\SetFigFont{12}{14.4}{\rmdefault}{\mddefault}{\updefault}$k$}}}
\put(3376,-2836){\makebox(0,0)[lb]{\smash{\SetFigFont{12}{14.4}{\rmdefault}{\mddefault}{\updefault}$r$}}}
\put(4426,-1411){\makebox(0,0)[lb]{\smash{\SetFigFont{12}{14.4}{\rmdefault}{\mddefault}{\updefault}$s$}}}
\end{picture}
\end{center}
\caption{Connected covering torus with $r=4$, $k=5$ and $s=3$.}
\label{fig1}
\end{figure}

The space on which coordinates which are arguments of the field in
eqn.(\ref{pq3}) take values is the torus depicted in fig. 1.  The
contribution to the path integral of this set of fields is denoted by
$ Z(\tau,\bar\tau)$ where $\tau=(s+i\nu r)/ k$.  It is the partition
function of a super-conformal field theory living on the torus
depicted in fig.1 and having target space $R^8$.  The boundary
conditions for the Bose fields are
\begin{eqnarray}
X^i(\sigma_1+k,\sigma_2)=X^i(\sigma_1,\sigma_2)\cr
X^i(\sigma_1,\sigma_2+r)=X^i(\sigma_1+s,\sigma_2)
\end{eqnarray}
and for the Fermi fields are
\begin{eqnarray}
\psi(\sigma_1+k,\sigma_2)=\psi(\sigma_1,\sigma_2)\cr
\psi(\sigma_1,\sigma_2+r)=(-1)^r\psi(\sigma_1+s,\sigma_2)
\end{eqnarray}
Note that the fermions are now not all anti-periodic.  When we move upward
by one of the small blocks in fig.1 they must be antiperiodic, thus the above
periodicity when we move up by $r$ blocks.

It is possible to change the coordinates so
that the integration region is the square torus $(\sigma_1, \sigma_2)\in
\left([0,1),[0,1)\right)$.  The necessary coordinate transformation is
\begin{eqnarray}
\sigma_1'=\frac{\sigma_1}{k} - \frac{s\sigma_2}{kr} ~~,~~
\sigma_2'= \frac{\sigma_2}{r}
\end{eqnarray}
Then the metric is 
\begin{equation}
g_{ab}=\left( \matrix{   1 & \tau_1\cr \tau_1& 
\vert\tau\vert^2\cr}\right)\ 
~~,~~\tau=\frac{s+i\nu r}{k}
\end{equation}
and the boundary conditions are
\begin{eqnarray}
X^i(\sigma_1+1,\sigma_2)=X^i(\sigma_1,\sigma_2)\cr
X^i(\sigma_1,\sigma_2+1)=X^i(\sigma_1,\sigma_2)\cr
\psi(\sigma_1+1,\sigma_2)=\psi(\sigma_1,\sigma_2)\cr
\psi(\sigma_1,\sigma_2+1)=(-1)^r\psi(\sigma_1,\sigma_2)
\end{eqnarray}
Note that the boundary condition for the Fermi field still depends on $r$.
When $r$ is even, 
the fermion and boson have the same boundary conditions.  
These sectors are supersymmetric.  The mode expansion of both the fermions 
and bosons contain zero modes.  Functional integration over bosonic zero modes 
produces a factor of the infinite 
volume of $R^8$ and integration over the fermionic
zero mode produces a factor of zero.  If we were computing the Witten index, 
this product of infinity times zero, suitably regulated would yield the 
number of zero energy states of the supersymmetric theory.  However, here, 
we are computing an extensive thermodynamic variable - the free energy - 
from which we must extract a factor of the volume of the space in order to 
obtain the free energy
density.  In this case, sectors which contain fermion zero modes do not 
contribute.   
On the other hand, when $r$ is  odd, the fermions have 
anti-periodic boundary conditions - supersymmetry is broken by this 
boundary condition - and there are no fermion zero modes in the mode 
expansion. These sectors will survive and contribute to the partition 
function.  Thus,
\begin{equation}
Z(\tau,\bar\tau)=0~{\rm when}~r~{\rm is ~ even}
\label{zermod}
\end{equation}

When the partition function is re-written as a sum over decompositions
into cycles of the commuting permutations, 
\begin{eqnarray} 
Z=\sum^\infty_{N=0}\frac{1}{N!}
e^{-N\beta/\sqrt{2}R}\sum_{\stackrel {P,Q} {PQ=QP} }
\cdot
\nonumber \\
\cdot \sum_{\rm decomp.}\prod_{\rm cycles} Z(\tau,\bar\tau)\ ,
\label{Zpqr}
\end{eqnarray}  
The sum over decompositions into cycles exponentiates to
the exponential of a sum over connected parts.  This was shown
explicitly in \cite{Grignani:1999sp} and also happens for a related
expression in ref.\cite{Billo:1999xc}.  In the end, the free energy of
the matrix string theory is given by the expression
\begin{equation}
F=-\frac{1}{\beta}
\sum_{N=0}^{\infty}
\sum_{\stackrel{kr=N}{r\ odd}}\sum_{s=0}^{N/r-1}
\frac{e^{-\frac{N\beta}{\sqrt{2}R}}}{N}
Z\left(\tau,\bar\tau\right)
\label{result}
\end{equation}
Note that $r$ gives the 
height of the torus, in the $\sigma_2$ direction and, taking into account 
eqn. ({\ref{zermod}) and the discussion before it, $r$ must be odd. 

With appropriate re-labeling of the integers, and evaluating the
functional integral $Z(\tau,\bar\tau)$, the expression (\ref{result}) is
identical to (\ref{pf4}), the expression for the free energy of the
DLCQ type II superstring.  Note that, in comparing this with
eqn.(\ref{pf4}), rather than $1/k^2$ appearing in the summand in
eqn.(\ref{result}), there is $1/N=1/kr$ where we label $k=N/r$.  This
difference of a factor of $k/r$ is absorbed in a factor of $1/\tau_2$
in eqn.(\ref{pf4}).  

\section{Remarks}

Given the good agreement of the genus 1 string partition function with
the Dijkgraaf-Verlinde-Verlinde limit of the matrix string model, it
would be interesting to check it a higher genus.  A concrete question
which could be asked is whether the two-point correlator of the twist
operator which ref.\cite{Dijkgraaf:1997vv} suggested was responsible
for interactions in the matrix string model could reproduce the genus
2 type II string partition function with (\ref{mod}) inserted.  In
fact, it has been found that, in the large $N$ limit, the correlation
function of twist operators on a symmetric orbifold indeed correctly
reproduces four-string amplitudes
\cite{Arutyunov:1998eq}-\cite{Arutyunov:1997gt}.

In addition, other tests of the matrix model are possible.  In
ref.\cite{Grignani:2001hb} the analysis of the partition function was
carried out in a constant background B-field.  This is a solvable
deformation of closed string theory on Minkowski space.  It was found
that a constraint similar to (\ref{mod}) appeared in the string path
integral.  Also, the modification of the matrix string model due to
the presence of a constant $B$ field was found and the zero string
coupling limit of that model was again shown to have identical
thermodynamic partition function to the type II string theory.  It
would be interesting to test this idea with other solvable
deformations of string theory. Some good examples are string theories
with D-branes and electromagnetic fields (for a review see
ref.\cite{Ambjorn:2000yr}) where, some interesting behavior is
observed in the finite temperature case \cite{Ambjorn:1999gt}.

It is well known that string theories, because of their exponentially
increasing density of states, have a limiting upper temperature.
Interacting strings are thought to undergo a phase transition at that
temperature
\cite{Atick:1988si}\cite{Sathiapalan:1986db}\cite{Kogan:jd}.  It was
shown in ref.\cite{Grignani:1999sp} that the Hagedorn temperature is
the same in the DLCQ as it is in ordinary light-cone or covariant
quantization of the string.  It comes from a divergence in the sum
over $N$ in equations like eqn.(\ref{pf33}) or eqn.(\ref{freenergy4}).
It was found in ref.\cite{Grignani:2001ik} that when a constant
external $B$-field is switched on, the Hagedorn temperature also
depends on $B$ in a way which was puzzling - the processes of
de-compactifying the light-cone and switching on $B$ did not commute
with each other.  An explanation which was given there was the
non-extensive behavior of the Hagedorn transition.

Another interesting question to do with the Hagedorn transition is to ask
where it would come from in the matrix model.  The matrix model had
partition function
\begin{equation}
Z[\beta]=\sum_{N=0}^\infty e^{-\frac{\beta N}{\sqrt{2}R}}
\int [dAdXd\psi]~e^{-S_E[A,X^i,\psi]}
\end{equation}
and again, one expects that the only possible source of Hagedorn behavior is
in the potential divergence of the summation over $N$.  Since the partition
function in the above expression is a sum over Yang-Mills partition functions
the only possible source of divergence from the sum over $N$ is governed by
the large $N$ limit of the Yang-Mills partition function.  

Typically, this is bad news.  In the generic large N limit, the
Yang-Mills theory free energy goes like
$$
\beta F\sim -N^2\cdot numbers
$$
simply because there are $N^2$ degrees of freedom.  This means that
the partition sum typically diverges! One needs some very special
tuning of parameters (such as the Dijkgraaf-Verlinde-Verlinde limit of
$g_s\to 0$) if it is to converge.  Even in the limit which gets free matrix
strings, it goes like 
$$
\beta F_{\rm DVV}\sim -N\beta_H/R
$$
where $\beta_H$ is the inverse Hagedorn temperature and the partition
sum converges only if the temperature is low enough ($\beta>\beta_H$.

To what do we attribute the generic divergence?  One suspect might be
the nucleation of black holes which would destroy the thermal
ensemble.  Estimates of the $N$-dependence of black hole free energy
in matrix theory \cite{Banks:1997hz}-\cite{Banks:1997tn} show that it
typically grows faster than $N$ (and must be negative).

Finally, we note that, in ref.\cite{D'Adda:2001nf} they found a phase
transition in a discrete gauge theory.  They interpreted the gauge theory
as a statistical model of branched covers of a triangulated Riemann surface
on one hand and a random walk on the gauge group on the other hand.  It is
easy to see that that phase transition coincides with the Hagedorn phase
transition of our DLCQ thermal partition functions if we interpret them
as statistical models of branched covers. (The basic difference between this
models and those of ref.\cite{D'Adda:2001nf} is that the string models
have continuous rather than triangulated Riemann surfaces, so there are
continuous moduli to be integrated over to get the partition function).

\end{document}